\documentclass[prd, twocolumn, nofootinbib]{revtex4}

\usepackage{epsfig} 
\newcommand{\om}{\Omega_M}
\newcommand{\LCDM}{$\Lambda$CDM\ }
\newcommand{\ode}{\Omega_{\rm DE}}
\newcommand{\ok}{\Omega_{\rm K}}
\begin{document} 

\title{Exploring the Dark Energy Redshift Desert with 
the Sandage-Loeb Test} 

\author{Pier-Stefano Corasaniti$^1$, Dragan Huterer$^2$, Alessandro
Melchiorri$^3$} 
\affiliation{
$^1$LUTH, CNRS UMR 8102, Observatoire de Paris-Meudon, 5 Place Jules Janssen,
92195 Meudon Cedex, France\\ 
$^2$Kavli Institute for Cosmological Physics and Astronomy and Astrophysics Department, 
University of Chicago, Chicago, IL 60637\\ 
$^3$Dipartimento di Fisica e Sezione INFN, Universita' degli Studi di
Roma ``La Sapienza'', Ple Aldo Moro 5,00185, Rome, Italy}
\begin{abstract}

We study the prospects for constraining dark energy at very high redshift with
the Sandage-Loeb (SL) test -- a measurement of the evolution of cosmic redshift
obtained by taking quasar spectra at sufficiently separated epochs. This test
is unique in its coverage of the ``redshift desert'', corresponding roughly to
redshifts between 2 and 5, where other dark energy probes are unable to provide
useful information about the cosmic expansion history.  Extremely large
telescopes planned for construction in the near future, with ultra high
resolution spectrographs (such as the proposed CODEX), will indeed be able to
measure cosmic redshift variations of quasar Lyman-$\alpha$ absorption lines
over a period as short as ten years. We find that these measurements can
constrain non-standard and dynamical dark energy models with high significance
and in a redshift range not accessible with future dark energy surveys.  As the
cosmic signal increases linearly with time, measurements made over several
decades by a generation of patient cosmologists may provide definitive
constraints on the expansion history in the era that follows the dark ages but
precedes the time when standard candles and rulers come into existence.

\end{abstract} 


\date{\today}

\maketitle 

\section{Introduction}\label{intro}

Measurements of luminosity distance to Type Ia supernovae (SNe; \cite{SN}) in
combination with the location of the acoustic peaks in the Cosmic Microwave
Background (CMB) power spectrum \cite{CMBpeak}, as well as the scale of the
baryon acoustic oscillations (BAO) in the matter power spectrum \cite{SDSSbao}
provide an accurate determination of the geometry and matter/energy content of
the universe. These measurements are almost exclusively sensitive to the
cosmological parameters through a time integral of the Hubble parameter (or,
the expansion rate).  Although direct measurements of the Hubble parameter are
feasible, they are typically difficult. For instance, the BAO probe radial
modes are directly sensitive to $H(z)$ \cite{Hu_Haiman}, but require
exceedingly precise knowledge of individual galaxy redshifts. In addition, the
few proposals to directly measure the expansion rate all propose to determine
$H(z)$ at a few specific epochs: $z\lesssim 2$ (from, say, the BAO, or by
measuring the relative ages of passively evolving galaxies
\cite{Jimenez_Loeb}), $z\sim 1000$ (from the CMB \cite{Zahn_Zal}) and $z\sim
10^9$ (from the Big Bang Nucleosynthesis).  In particular, a new cosmological
window would open if we could directly measure the cosmic expansion within the
``redshift desert'', ideally exploring $2\lesssim z\lesssim 5$.

During the early years of Big-Bang cosmology Allan Sandage studied a
possibility of directly measuring the temporal variation of the redshift of
extra-galactic sources \cite{Sandage}. As explained in the next section, this temporal
variation is directly related to the expansion rate at the source redshift.
However, measurements performed at time intervals separated by less than $10^7$
years would have failed to detect the cosmic signal with the technology available at that time
\cite{Sandage}.  In 1998 these ideas have been revisited by Loeb \cite{Loeb}.
He argued that spectroscopic techniques developed for detecting the reflex
motion of stars induced by unseen orbiting planets could be used to detect the
redshift variation of quasar (QSO) Lyman-$\alpha$ absorption lines.  A sample
of a few hundred QSOs observed with high resolution spectroscopy with a $\sim
10$ meter telescope could in fact detect the cosmological redshift variation at
$\sim 1\sigma$ in a few decades. In what follows we will therefore refer to
this method as the ``Sandage-Loeb'' (SL) test.

The astronomical community has since entertained increasingly ambitious
ideas with proposals for building a new generation of extremely large
telescopes ($30-100$ meter diameter) \cite{GMT,TMT,Euro50,ELT}.
Equipped with high resolution spectrographs, these powerful machines could
provide spectacular advances in astrophysics and cosmology. The
Cosmic Dynamics Experiment (CODEX) spectrograph has been recently proposed to
achieve such a goal \cite{Pasquini,Molaro}.

The large number of absorption lines typical of the Lyman-$\alpha$
forest provide an ideal method for measuring the shift velocity. The latter can
be detected by subtracting the spectral templates of a quasar taken at two
different times. As quasar systems are now readily targeted and observable in the
redshift range $2\lesssim z\lesssim 5$, we have a new test of the 
cosmic expansion history during the epoch just past the dark ages
when the first objects in the universe are forming.

If dark energy is consistent to our simplest models --- small zero-point energy
of the vacuum or a slowly rolling scalar field --- then it significantly speeds
up the expansion rate of the universe at $z\lesssim 1$.  Under the same
assumption, dark energy is subdominant at redshift $z\gtrsim 1$, and almost
completely negligible at $z\gtrsim 2$.  One may therefore ask whether is
worthwhile to probe $z\gtrsim 2$ anyway. The answer is affirmative: since we do
not know much about the physical provenance of dark energy, it is useful to
adopt an entirely empirical approach and look for the signatures of dark energy
at all available epochs regardless of the expectations. This question has
recently been studied in some detail by Linder \cite{Linder_darkages}, who
considered toy models of dark energy that have non-negligible energy density at
high redshift.

In this paper we study the cosmological constraints that can be inferred from
future observations of velocity shift and their impact on different classes of
dark energy models. In Sec.~\ref{sec:SL} we review the physics behind the SL
test, in Secs.~\ref{sec:standard} and \ref{sec:nonstandard} we describe future
constraints on standard and non-standard dark energy models respectively, and
in Sec.~\ref{sec:disc} we discuss our results and future prospects.

\section{The Sandage-Loeb test} \label{sec:SL}

It is useful to firstly review a standard textbook calculation of the
Friedmann-Robertson-Walker cosmology. Consider an isotropic source emitting at
rest. Since it does not posses any peculiar motion, the comoving distance to an
observer at the origin remains fixed. Then waves emitted during the
time interval ($t_{\rm s}$, $t_s+\delta{t}_{\rm s}$) and detected later during
($t_o$, $t_o+\delta{t}_o$) satisfy the relation \cite{Sandage}
\begin{equation}
\int_{t_{\rm s}}^{t_o}\frac{dt}{a(t)}=\int_{t_{\rm s}+
  \delta{t}_{\rm s}}^{t_o+\delta{t}_o}\frac{dt}{a(t)}.\label{fr}
\end{equation}
where $t_{\rm  s}$ is the time of emission and $t_o$ the time of observation. 
For small time intervals ($\delta{t}\ll t$) this gives the well known redshift relation
of the radiation emitted by a source at $t_{\rm s}$ and observed at $t_o$, 
\begin{equation}
1+z_{\rm s}(t_o)=a(t_o)/a(t_{\rm s}).
\end{equation}
Let us now consider waves emitted after a period $\Delta{t}_{\rm s}$ at $t_{\rm
s}+\Delta{t}_{\rm s}$ and detected later at $t_o+\Delta{t}_o$.  Similarly to
the previous derivation the observed redshift of the source at
$t_o+\Delta{t}_o$ is
\begin{equation}
1+z_{\rm s}(t_o+\Delta{t}_o)=a(t_o+\Delta{t}_o)/a(t_{\rm s}+\Delta{t}_{\rm s}).
\end{equation}
Therefore an observer taking measurements at times $t_o$ and $t_o+\Delta{t}_o$
would measure the following variation of the source redshift
\begin{equation}
\Delta{z}_s\equiv\frac{a(t_o+\Delta{t}_o)}{a(t_{\rm s}+\Delta{t}_{\rm s})}-
\frac{a(t_o)}{a(t_{\rm s})}.\label{dz}
\end{equation}
In the approximation $\Delta t/t\ll 1$, we can expand
the ratio $a(t_o+\Delta t_o)/a(t_{\rm s}+\Delta t_{\rm s})$ to linear order
and further using the relation $\Delta{t}_o=[a(t_o)/a(t_{\rm s})]\Delta{t}_{\rm
s}$ (as it can be easily inferred from Eq.~(\ref{fr})) we obtain
\begin{equation}
\Delta z_s \approx \left[\frac{{\dot a}(t_{\rm o})-{\dot a}(t_{\rm s})}{a(t_{\rm s})}\right]\Delta{t}_o.
\end{equation}
This redshift variation can
be expressed as a spectroscopic velocity shift, $\Delta v\equiv c\Delta z_s
/(1+z_{\rm s})$. Using the Friedmann equation to relate ${\dot a}$ to the
matter and energy content of the Universe we finally obtain (see \cite{Loeb}):

\begin{equation}
\frac{\Delta{v}}{c}=H_0\Delta{t}_0\left [1-{E(z_s)\over 1+z_s}\right ],
\label{dv}
\end{equation}

\noindent where $H_0$ is Hubble constant, $E(z)\equiv H(z)/H_0$ is the (scaled) 
Hubble parameter at redshift $z$, $c$ is the speed of light, and
we have normalized the scale factor to $a(t_o)=1$ and
neglected the contribution from relativistic components. For a
constant dark energy equation of state we have

\begin{equation}
E(z) = \left [\Omega_M(1+z)^3+\Omega_{DE}(1+z)^{3(1+w)}+\Omega_K(1+z)^2\right ]^\frac{1}{2}
\end{equation}

\noindent where $\Omega_M$ and $\Omega_{DE}$ are the matter and dark energy density
relative to critical, $\Omega_K=1-\Omega_M-\Omega_{DE}$ is the curvature, and $w$ is the
dark energy equation of state.

\begin{figure}[t]
\begin{center} 
\psfig{file=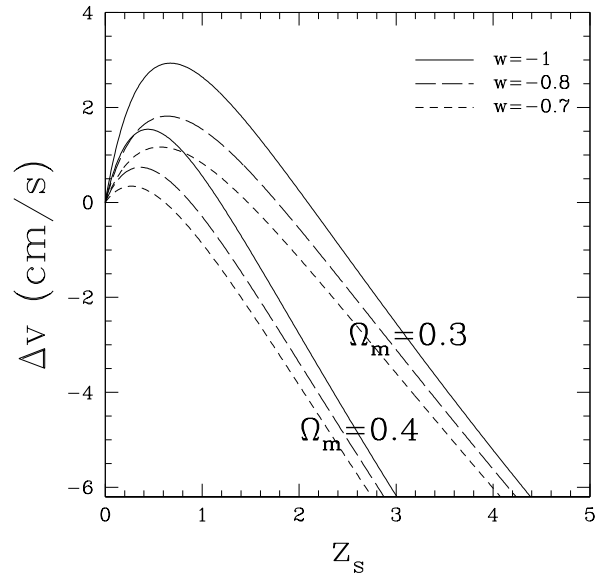, height=3.2in}
\caption{Cosmic velocity shift as function of the source redshift in a flat
universe for different values of $\Omega_M$ and $w$. A time interval of $10$
years has been assumed.  The signal primarily depends on $\Omega_M$ and more weakly
on $w$.  }
\label{fig:1} 
\end{center} 
\end{figure}

In Figure~\ref{fig:1} we plot $\Delta v$ as function of the source redshift in
the flat case for different values of $\Omega_M$ and $w$ assuming a time
interval $\Delta t_{\rm o} = 10$ years.  As we can see $\Delta v$ is positive
at small redshifts and becomes negative at 
$z\gtrsim 2$.  Under the assumption
of flatness the amplitude and slope of the signal depend mainly on
$\Omega_{M}$, while the dependence on $w$ is weaker.

In spite of the tiny amplitude of the velocity shift, the absorption lines in
the quasar Ly-$\alpha$ provide a powerful tool to detect such a small
signal. As already pointed out in \cite{Loeb} the width of these lines is of
order $\sim 20$ ${\rm km/s}$, with metal lines even narrower. Although these
are still a few orders of magnitude larger than the cosmic signal we seek to
measure, each Ly-$\alpha$ spectrum has hundreds of lines.  Therefore
spectroscopic measurements with a resolution $R>20000$ for a sample of $\sim
100$ QSOs observed $\sim 10$ years apart can lead to a positive detection of
the cosmic signal. Moreover astrophysical systematic effects such as peculiar
velocities and accelerations can lead to negligible corrections \cite{Loeb}.
Local accelerations may indeed be more important, but due to their direction
dependence they can be determined from velocity shift
measurements of QSOs sampled in different directions on the sky.

In \cite{Pasquini} the authors have performed Monte Carlo simulations of
Lyman-$\alpha$ absorption lines to estimate the uncertainty on $\Delta v$ as 
measured by the CODEX spectrograph. The statistical error can be parametrized in
terms of the spectral signal-to-noise $S/N$, the number of Ly-$\alpha$ quasar systems
$N_{QSO}$, and the quasar's redshift

\begin{equation}
\sigma_{\Delta v} = 1.4 \left(\frac{2350}{S/N}\right)\sqrt{\frac{30}{N_{QSO}}}
\left(\frac{1+z_{QSO}}{5}\right)^{-1.8} {\rm \frac{cm}{s}},\label{sig}
\end{equation}

\noindent (at $z<4$) where $S/N$ is defined per $0.0125A$ pixel. 
The source redshift dependence becomes flat at $z>4$. The numerical factor
slightly changes with the source redshift, varying from $1.4$ at $z=4$ to
$2$ at $z\approx 2$. This small variation arises because the number of observed absorption lines decreases 
with $z_{QSO}$. For simplicity we assume its value to be fixed to $1.4$. 
The large $S/N$ necessary to detect the cosmic signal implies that a positive
detection is not feasible with current telescopes. However the CODEX
spectrograph is currently being designed to be installed on the ESO Extremely Large
Telescope; a $\sim 50$ meter giant that can reach the necessary
signal-to-noise with just few hours integration.

In the next sections we describe the cosmological window that such
observations could open, with particular focus on dark energy models.

\begin{figure}[!t]
\begin{center} 
\psfig{file=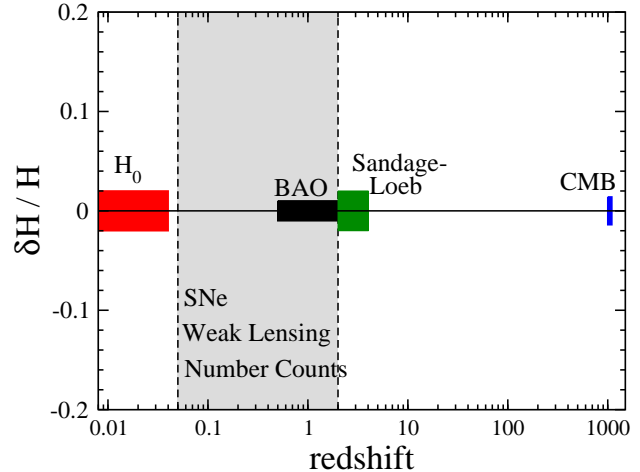, height=3.5in, angle=-90} 
\caption{Fractional accuracy in the measurement of the Hubble parameter $H(z)$
as a function of redshift for a sample of cosmological probes.  The accuracy
and redshift ranges shown are best-guess values for the future surveys, and
assume a {\it single}  measurement of the Hubble parameter for each probe.
Since SNe, number counts of clusters, and weak gravitational
lensing do not measure the Hubble constant directly, but rather 
some combination of distances and (in the case of the latter two) growth
of density perturbations, we only indicate their redshift range with
the shaded region. The Sandage-Loeb overall constraint assumes roughly a 30 year
survey and other specifications as in the text.  
}
\label{fig:2} 
\end{center} 
\end{figure}

\section{Cosmological Parameters and Standard Dark Energy Models}\label{sec:standard}

We forecast constraints on cosmological parameters from velocity shift
measurements using the Fisher matrix method for a LCDM fiducial cosmology. We
assume experimental configuration and uncertainties similar to those expected
from CODEX \cite{Pasquini,Molaro}. Namely we consider a survey observing a total of $240$
QSOs uniformly distributed in $6$ equally spaced redshift bins in the 
range $2\lesssim z \lesssim 5$, with a signal-to-noise $S/N=3000$, and the
expected uncertainty as given by Eq.~(\ref{sig}). Since there is no time
integration involved in the computation of $\Delta v$ the Fisher matrix
components can be easily determined analytically.

In Figure~\ref{fig:2} we show the near-future status of measurements of the
Hubble parameter at different cosmological epochs. We assume upcoming
measurements of $H_0$ accurate to 2\%, the overall BAO measurements of the expansion
rate to 1\% \cite{DETF}, the CMB measurement of 1.4\% \cite{Zahn_Zal}, and the
SL test measurement of $H$ using the assumptions outlined above. For
clarity we do not show the Big Bang Nucleosynthesis constraint on $H(z)$ with
$z=z_{\rm bbn}\approx 10^9$, which is accurate to $\sim 10$\% (e.g.\
\cite{Copi}) and will improve as soon as the baryon density and deuterium
abundance are more accurately determined. Since Type Ia supernovae, number
counts of clusters, and weak gravitational lensing do not measure the Hubble
constant directly, we only indicate their approximate redshift range with the shaded
region. 
Clearly, the SL test is probing an era not covered with any other reliable
cosmological probe. 

\begin{figure}[!t]
\begin{center} 
\psfig{file=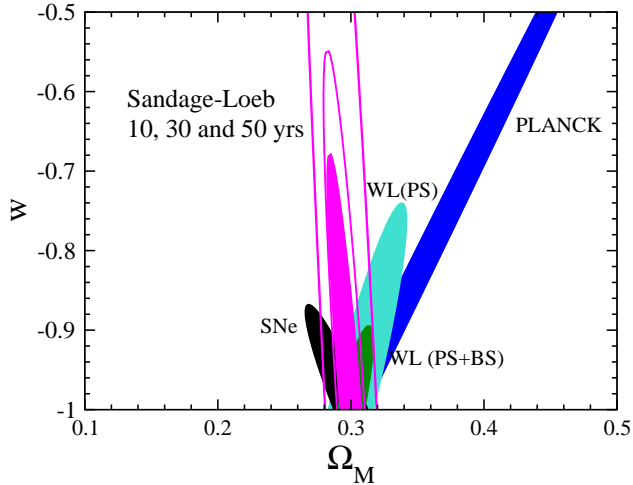, height=3.5in, angle=-90} 
\caption{Constraints in the $\om-w$ plane for the SL test assuming (with
increasingly smaller contours) a 10, 30 and 50-year survey and the number of
quasars as in the text. We also show constraints expected from type Ia supernovae,
weak lensing power spectrum, power spectrum in combination with bispectrum,
and constraints expected from Planck's measurement of distance to the last
scattering surface. SNa and weak lensing estimates are based on the SNAP mission,
and SNe also include systematic errors. The SL test contour is 
complementary to other constraints, and is mostly sensitive to the matter
density, represented here by $\om$. 
}
\label{fig:3} 
\end{center} 
\end{figure}

In Figure~\ref{fig:3} we plot the $1\sigma$ contours in the $\Omega_M-w$ plane
as expected from type Ia supernovae, weak lensing power spectrum, power
spectrum plus the bispectrum, and constraints expected from Planck's
measurement of distance to the last scattering surface. Supernova and weak
lensing estimates are based on the SNAP mission \cite{SNAP}, and SNe include
systematic errors. The SL contour are complementary to those of other tests
since $\Delta v$ probes a different degeneracy line in the $\Omega_M-w$
plane. It is worth noticing that such measurements are mostly sensitive to the
matter density, as expected from the plot shown in Fig.~\ref{fig:1}. Allowing
also for variation of the curvature, $\ok$, we find that the SL test alone
determines the matter density about four times better than the curvature. As an
example, for the 30-year survey the marginalized errors are $\sigma(\om)=0.03$ and
$\sigma(\ok)=0.13$.

We have also found that limited accuracy in the Hubble constant measurement does
not degrade the power of the SL test. For example, if $h$ is known to 0.04 (or
to about 5\%), the accuracy in $w$ is degraded by only 2\% relative to the case
when $h$ is perfectly known.

As far as dark energy is concerned, assuming a Gaussian prior on $h$ with
$\sigma_h=0.04$, we obtain $\sigma_w=0.8$ for $\Delta{t}_0=10$ years time
interval and $0.3$ for $30$ years. Thus the constraints on $w$ are not
competitive with those inferred from other, well-established tests such as SNe
Ia, weak lensing or BAO once we take into account that the latter probes will
provide strong constraints by the time the SL test is undertaken. However, one
should note that the constraints obtained by SL decrease {\it linearly} with
time.  For measurements made over a century, and with the expectedly larger
number of QSOs, the SL limits on $w$ can easily be at the few percent level.
 
\section{Non-standard dark energy models}\label{sec:nonstandard}

A unique advantage of the SL test is that it probes the redshift range
$2\lesssim z\lesssim 5$ which is very difficult to access otherwise.  As
mentioned in Section~\ref{intro}, probing this redshift range is important for
testing non-standard dark energy models that would otherwise be
indistinguishable from those with a smooth, nearly or exactly
constant equation of state function $w(z)$. 

\begin{figure}[t]
\psfig{file=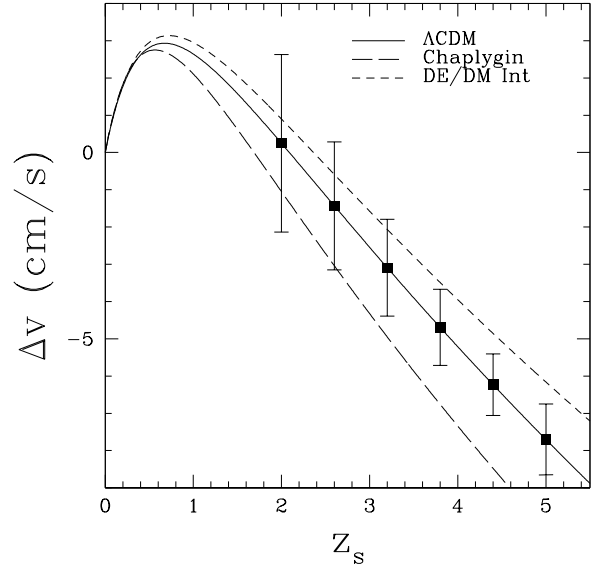, height=3.2in}
\caption{Cosmic velocity shift as function of the source redshift for a flat
\LCDM model with $\Omega_m=0.3$ (solid line), Chaplygin gas model (long dashed
line) and an interactive dark energy-dark matter model (short dashed line) with
$2\%$ deviation from standard redshift scaling of matter (see text).  The error
bars on the \LCDM model assume a 10 year survey and other specifications as in
the text.  }
\label{fig:4}
\end{figure} 

WMAP observations in combination with low-redshift limits from SNe Ia impose a
weak upper bound on the amount of dark energy deep during matter domination to be
$\Omega_{DE}(z)<0.1$ \cite{Caldwell03,Corasaniti04}. Let us suppose 
that dark energy re-emerged at $3<z<5$ with $\ode(z)=0.1$ in
this range, while essentially zero at larger redshifts and that it behaved as a
standard \LCDM at $z<3$ (i.e.\ assuming standard $\ode=0.75$, $w=-1$
values). We would like to distinguish such a model from a pure \LCDM with $\ode$ and
$w$ as above. Clearly, low-redshift probes (SNe, BAO, galaxy clustering)
cannot distinguish these two models since the required redshift is too high even
for the most ambitious surveys. Furthermore, the distance to the last scattering
surface between the two models differs only by 0.5\%, which is too low to be
observable even with Planck, since the 1-$\sigma$ uncertainty is about 0.4\% with
temperature and polarization information \cite{EisHuTeg}. However, the two
models {\it can} be distinguished via the SL test at about 3-$\sigma$ level,
assuming only a 10-year survey and other specifications as in the previous section.

While the aforementioned scenario with dark energy emerging in the specific
window at $3\lesssim z\lesssim 5$ may seem contrived, it is easy to find
physically motivated models whose identification can significantly benefit from
data in this ``desert''.  One example is given by scalar field models which predict 
the periodic emergence of DE at various epochs during the history of the universe
\cite{oscillating,Griest}. Even though for these particular models one has to
go to a much higher redshift to see the next phase where $\ode=O(1)$, it is
plausible, and certainly currently observationally allowed (e.g.\
\cite{flowroll}) that such a phase could have occurred somewhere within
$2\lesssim z\lesssim 5$. 

Another example is given by models with dark energy-dark matter interaction
(see e.g. \cite{Damour,Amendola,Chimento,KhouryWelt,Peebles,Das,Alimi}).  In
the simplest realization where the scalar field only couples to dark matter, it
mediates a long range interaction which causes two separate effects.  First,
dark matter particles, unlike the baryons, experience a scalar-tensor type of
gravity, which modifies the Newtonian regime (see \cite{AmendolaPert}).  The
time and scale of when such type of modification becomes cosmologically
relevant depend on the particular model considered.  Second, dark matter
particles acquire a time dependent mass whose evolution is determined by the
specifics of the scalar field dynamics. As a consequence of this, the redshift
evolution of the dark matter density deviates from the usual $(1+z)^3$. At low
redshift, when the universe is dark energy dominated, these models cannot be
distinguished from the standard \LCDM. Therefore, the best way to probe models
with such dark matter-dark energy interaction is to map out cosmic expansion
during the matter dominated phase (see Figure~\ref{fig:4}).  The SL tests
offers a unique tool to do just that.

In order to forecast how well a deviation of the dark matter density from the
$(1+z)^3$ law can be detected, we parametrize its redshift evolution as
$\propto (1+z)^{3(1-b)}$ in the range $2\lesssim z \lesssim 5$, where $|b|<1$
is a constant free parameter. The scalar field, on the other hand, can be treated
as a dark energy component with $w\approx -1$, since the field slowly rolls
toward the minimum of its effective potential at late times \cite{Das}.  Assuming a
flat fiducial model with $\Omega_M=0.3$ and $w=-1$, we find that the SL test
can detect deviations from the standard matter scaling as small as $1\%$
(i.e. $b=0.01$) over $10$ years and $0.3\%$ for $30$ years. 
Therefore, SL test can provide constraints an order of magnitude tighter than
those inferred using future SNe Ia or the Alcock-Paczynsky test
\cite{Dalal}.  Since the deviation $b$ is generally a function of redshift, one
can use the velocity shift measurements to reconstruct the
redshift dependence of $b$, and then determine the strength and functional form
of the scalar interaction.

The Chaplygin gas is yet another dark energy candidate that can be tested in
the range of redshift probed by the SL test. Proposed as a phenomenological
prototype of unified dark energy and dark matter model \cite{kam,bil,ben}, it
describes an exotic fluid with an inverse power law homogeneous equation of
state, $ P=-|w_{0}|\Omega_{\rm Ch}^0\rho_0/\rho^{\alpha}$ (e.g.\ \cite{Bean_Dore}), where $w_{0}$ is the
present equation of state, $\Omega_{\rm Ch}^0$ is the current energy density of
the gas $\rho_0=3H_0^2$ is the total energy density and $\alpha >-1$ is a
dimensionless parameter. This corresponds to a fluid which behaves as dust in
the past and as cosmological constant in the future. For $\alpha=0$ the model
reduces to $\Lambda$CDM \cite{ave03}.  The Chaplygin gas energy density evolves
with redshift according to:

\begin{equation}
\Omega_{\rm Ch}(z)=\Omega_{\rm Ch}^0\left [-w_0+(1+w_0)
(1+z)^{3(\alpha +1)}\right ]^{1\over 1+\alpha }.
\end{equation}

As shown in \cite{Amendolafin} this
model can provide a good fit to current cosmological observables with
$\Omega_{\rm Ch}\sim 0.95$ (with a baryonic component of $\Omega_b=0.05$),
$w_{0}\sim -0.75$ and $\alpha=0.2$. From Fig.~\ref{fig:4} we can see that
although this model has a velocity shift at $z < 1$ similar to that of a \LCDM,
it can be tested with high redshift measurements. For instance  
we find that, for this particular model, the Chaplygin parameters can be
determined with uncertainties $\sigma_{w_0}=0.03$ and $\sigma_{\alpha}=0.04$
respectively and thus distinguished from the \LCDM values at a high
confidence level.

\section{Discussion}\label{sec:disc}

In this paper we have analyzed the prospects for constraining dark energy at
high redshift ($2\lesssim z\lesssim 5$) by direct measurements of the temporal
shift of the quasar Lyman-$\alpha$ absorption lines (the Sandage-Loeb
effect). While the signal is extremely small, the physics is straightforward,
and the measurement is certainly within reach of future large telescopes with
high resolution spectrographs. 


As the SL test mostly probes the matter density at high redshift, the
constraints on standard dark energy models with a nearly or exactly constant
equation of state $w$ are weaker than those that observations of SN Ia, BAO,
weak lensing and number counts will be able to achieve in the future (although
the SL test becomes competitive for measurements spread over a period of
several decades or more). This is mostly because the sensitivity of
cosmological probes to standard dark energy models is exhausted at $z\lesssim
2$ and higher redshift data do not improve them significantly (e.g.\ \cite{frieman}).
However, the principal power of SL measurements at $2\lesssim z\lesssim 5$
comes from their ability to constrain {\it non-standard} dark energy models,
where the dark energy density is non-negligible at higher redshifts --- or
equivalently, models where the total energy density does not scale with
redshift as $(1+z)^3$ at $z\gtrsim 2$. In particular, as discussed in the previous section, 
in only one decade the SL measurements
will allow to test the redshift scaling of the matter density with an accuracy
one order of magnitude greater than standard cosmological tests.

Tighter constraints on standard dark energy models could be obtained if observations 
of Ly-$\alpha$ systems were feasible at $z\lesssim 2$. What are the prospects for performing 
the SL test at lower redshift? For this to be possible UV space-based instruments are necessary.
Space-based UV Lyman-$\alpha$ astronomy is possible and has already produced
remarkable results (see e.g \cite{penton04}), though it generally lacks the
spectral resolution and wavelength coverage of the higher redshift
studies. However during the past few years high quality observations of several low
redshift QSOs have been obtained with the Hubble Space Telescope (HST) and
its Space Telescope Imaging Spectrograph (STIS).  It is therefore conceivable
that future space based experiments will be able to measure the SL effect at low redshift.

Other astrophysical probes that can potentially explore the redshift desert are not yet
well understood. For instance, it might be possible to measure the angular diameter
distance at redshift of 10-20 from the acoustic oscillations in the power-spectrum
of the 21cm brightness fluctuations \cite{Barkana_Loeb}.
Further, gamma ray bursts (GRBs) (e.g.\ \cite{GRB}) have been proposed 
as alternative standard candles. These can probe roughly the same redshift range as 
the SL ($z\lesssim 6$). Similarly gravitational wave ``standard sirens'' can measure distances to GW
sources all the way out to $z\sim 20$ \cite{Holz_Hughes}. However, the
prospects of turning the GRBs into standard candles is extremely uncertain at
this time, and it is not clear that they can be used as cosmological
probes. The GW sirens, while potentially providing amazingly accurate distance
measurement, suffer from gravitational lensing of the signal, which adds an
effective noise term to distance measurements and greatly degrades their power
to constrain cosmological models \cite{Holz_Hughes}. Conversely, the SL test is
based on extremely simple physics and involves controllable systematic errors,
but it does require a powerful instrument and patience to wait at least a
decade before repeating the measurements in order to produce interesting results.

We finally point out that the velocity shift signal increases linearly with time, thus
amply rewarding increased temporal separation between measurements. Therefore 
observations made over a period of several decades by a generation of 
patient cosmologists may provide definitive
constraints on the expansion history in the era before the usual standard
candles and rulers, Type Ia supernovae and acoustic oscillations in the
distribution of galaxies, become readily available.

\section*{Acknowledgments} 

It is a pleasure to thank Paolo Molaro and Matteo Viel for useful
discussions, and Avi Loeb for comments on the manuscript.
PSC is supported by CNRS with contract No. 06/311/SG.  DH has been supported by
NSF Astronomy and Astrophysics Postdoctoral Fellowship under Grant No.\
0401066.

\end{document}